\documentclass[conference]{IEEEtran}
\IEEEoverridecommandlockouts
\usepackage{cite}
\usepackage{amsmath,amssymb,amsfonts}
\usepackage{algorithmic}
\usepackage{graphicx}
\usepackage{textcomp}
\usepackage{xcolor}
\usepackage{tikz}
\usepackage{svg}
\usepackage{pgfplots}
\pgfplotsset{compat=newest}
\pgfplotsset{plot coordinates/math parser=false}
\newlength\figureheight
\newlength\figurewidth
\newcommand{\cb}{\textcolor{black}}

\def\BibTeX{{\rm B\kern-.05em{\sc i\kern-.025em b}\kern-.08em
    T\kern-.1667em\lower.7ex\hbox{E}\kern-.125emX}}
\begin{document}

\title{Mutli-Level Autoencoder: Deep Learning Based Channel Coding and Modulation 
%{\footnotesize %\textsuperscript{*}Note: Sub-titles are not captured in Xplore and
}

\author{\IEEEauthorblockN{Ahmad Abdel-Qader}
\IEEEauthorblockA{Department of Computer Science \\
University of British Columbia\\
Kelowna, Canada \\
aabdelqa@student.ubc.ca}
\and
\IEEEauthorblockN{Anas Chaaban}
\IEEEauthorblockA{School of Engineering \\
University of British Columbia\\
Kelowna, Canada \\
anas.chaaban@ubc.ca}
\and
\IEEEauthorblockN{Mohamed S. Shehata}
\IEEEauthorblockA{Department of Computer Science \\
University of British Columbia\\
Kelowna, Canada\\
mohamed.sami.shehata@ubc.ca}
}
\maketitle

\begin{abstract}
In this paper, we design a deep learning-based convolutional autoencoder for channel coding and modulation. The objective is to develop an adaptive scheme capable of operating at various signal-to-noise ratios (SNR)s without the need for re-training. Additionally, the proposed framework allows validation by testing \emph{all} possible codes in the codebook, as opposed to previous AI-based encoder/decoder frameworks which relied on testing only a small subset of the available codes. This limitation in earlier methods often led to unreliable conclusions when generalized to larger codebooks. In contrast \cb{to previous methods}, our multi-level encoding and decoding approach splits the message \cb{into blocks}, where each encoder block processes a distinct group of \cb{$B$ bits}. By doing so, the proposed scheme can exhaustively test \cb{$2^{B}$ }possible codewords for each encoder/decoder \cb{level, constituting a layer of the overall scheme. The} proposed model was compared to classical polar codes and TurboAE-MOD schemes, showing improved reliability with achieving comparable, or even superior results in some settings. Notably, the architecture can adapt to different SNRs by selectively removing one of the \cb{encoder/decoder layers} without re-training, thus demonstrating flexibility and efficiency in practical wireless communication scenarios.
\end{abstract}

\section{Introduction}

Ensuring reliable and low-latency transmissions has become a key priority for emerging wireless networks, especially given the rise of time-critical applications such as vehicle-to-vehicle/infrastructure communications, augmented/virtual reality, and industrial automation \cite{intro_1}. Meeting these strict requirements often relies on short packet communications implemented via finite blocklength (FBL) coding techniques \cite{intro_2}. In contrast to the infinite blocklength regime, where transmission rates can approach Shannon’s capacity with negligible error, FBL scenarios inherently exhibit non-zero error probabilities that become more pronounced at shorter blocklengths. Pioneering work by Polyanskiy et al. \cite{Polyanskiy} confirmed these effects by deriving a closed-form expression for decoding error probability under additive white Gaussian noise (AWGN) channels, illustrating how reliability deteriorates in the short-packet regime even when operating below Shannon capacity \cite{shannon}.

Traditionally, channel coding and modulation have been \cb{handled either as disjoint modules predominantly for simplicity of implementation and analysis, or combined in a coded-modulation to improve performance. N}ew use cases within the FBL regime require tightly integrated and highly flexible methods \cite{intro_3}. Although fifth generation (5G) standards have adopted polar and LDPC codes for moderate to long blocklengths \cite{intro_4}, \cite{intro_5}, reliably operating at extremely short code lengths remains challenging, prompting the exploration of novel designs.

An emerging trend aims to unify coding and modulation to enhance efficiency and adaptability. Multilevel polar coded modulation (MLPCM) is one such approach, offering measurable gains over standard polar codes in Gaussian channels \cite{intro_6}. Nonetheless, these techniques may underperform when blocklengths become very short or when latency constraints are especially tight. In response, neural network (NN)-based strategies have gained traction due to their potential to learn end-to-end communication mappings at near optimal performance with lower complexity than most of conventional channel codes \cite{intro_7}, \cite{intro_8}.

Initial neural networks based methods replaced select physical-layer components \cite{intro_9}, including neural decoders \cite{intro_10} and neural modulators/demodulators \cite{intro_11}. Further research then evolved into fully end-to-end (E2E) autoencoder (AE) communication systems \cite{intro_12, intro_13, intro_14,intro_15}. Autoencoders, in general, excel at learning internal latent features by minimizing reconstruction errors and have been deployed for tasks like denoising \cite{intro_16} or generative modeling \cite{intro_17}. While under-complete AEs learn compressed latent representations, which can help in channel compression, over-complete AEs expand the input dimensionality to facilitate more robust signal recovery. This over-complete property is well-aligned with the need to map a limited set of input bits to a higher-dimensional codeword that can withstand noise. More advanced deep learning–based channel coding has garnered attention for its ability to jointly learn encoder and decoder structures. The work in \cite{turboAE} introduces Turbo Autoencoder (TurboAE), a fully end-to-end scheme consists of a neural network based channel encoder convolutional neural networks (CNN) with an interleaver and a neural network based decoder that achieves lower error rate at different SNRs as compared to multiple traditional codes under moderate block lengths. In \cite{turboAE-mod}, the authors extend TurboAE to jointly learn both coding and modulation (TurboAE-MOD) using a feed-forward neural network (FFNN) for modulation, demonstrating performance comparable to modern coded-modulation schemes under various channel settings and showing the potential of learned modulations. Going further, \cite{cnn_autoenc} investigates a CNN-Autoencoder architecture for finite blocklength Gaussian channels, illustrating that it can outperform conventional polar and Reed-Muller–based coded modulation while approaching the theoretical maximum achievable rate.

Despite these promising advances, a key drawback encountered in many deep learning-based channel coding frameworks is the reliance on a small subset of the full codebook for training and testing. While this reduces the computational burden, it can mask potential performance degradation when generalized to unseen codes from the codebook or different SNR regimes. 

Inspired by multi-level codes \cite{intro_18}, we introduce a multi-level autoencoder (MLAE) that addresses these issues by splitting each message into \cb{$B$-bit blocks, ensuring exhaustive coverage of $2^{B}$ possible codewords per level when training or testing.} By systematically evaluating all feasible codewords, we can better assess reliability across the entire codebook. This defines a vital capability often neglected in existing methods that rely on partial sampling for computational convenience. Moreover, the proposed scheme is designed to adapt across varying SNR conditions without the need for repeated re-training. In practical wireless systems where SNR may fluctuate due to mobility or channel fading, this adaptability is crucial to maintaining consistent performance. \cb{Additionally, our approach centers on jointly designing the encoder/modulator and demodulator/decoder structures to achieve robust performance in short code-lengths. Overall, this paper aims to:}
\begin{itemize}
    \item Demonstrate the potential of a multi-level convolutional autoencoder design that exhaustively evaluates the entire codebook for \cb{FBL.}
    \item Address the growing need for adaptive, low-latency schemes capable of supporting a wide range of SNR conditions without re-training.
    %\item Compare the proposed model against well-known channel coding techniques (polar codes, LDPC codes, and Turbo-AE), highlighting its superior performance in the FBR.
    \item Contribute to the broader research goal of learning channel codes that can effectively operate under non-asymptotic conditions.
\end{itemize}

The rest of this paper is organized as follows. Section~II describes the system model. Section~III presents the proposed MLAE design and training methodology. Numerical results and discussions are provided in Section~IV. Finally, Section~V concludes the paper and outlines future research directions.

\section{System Model}

Consider a communication system where a transmitter encodes a message \(m\), composed of \(K\) information bits, into a complex codeword \(\mathbf{x} = (x_1, x_2, \dots, x_n) \in \mathbb{C}^n\). The codeword must satisfy the average power constraint
\[
\frac{1}{n} \sum_{i=1}^{n} |x_i|^2 \le P.
\]
Once transmitted over a Gaussian channel, the received vector is
\[
\mathbf{y} = \mathbf{x} + \mathbf{w},
\]
where \(\mathbf{w}\) is an \(n\)-dimensional vector of i.i.d.\ circularly symmetric complex Gaussian noise with zero mean and variance \(N_0\). On the receiver side, the goal is to produce an estimate \cb{\(\hat{\mathbf{m}}\) of the original message \(\mathbf{m}\) after receiving $\mathbf{y}$} using the channel conditional probability \(P_{\mathbf{Y}|\mathbf{X}}(\mathbf{y}|\mathbf{x})\). The design requires that the frame error probability (FEP) \cb{\(\Pr(\hat{\mathbf{m}} \neq \mathbf{m})\)} remain below a prescribed threshold \(\varepsilon\). The interplay between the blocklength \(n\) (which influences latency) and \(\varepsilon\) (which dictates reliability) constrains the achievable information rate of the system.

\medskip

In a \emph{conventional} architecture, this encoding process typically occurs in two stages. First, a channel encoder maps the \(K\) input bits into a length-\(N\) binary codeword at a coding rate
\[
R_{\text{cod}} = \frac{K}{N}.
\]
An \(M\)-ary modulator of order \(M = 2^{k_{\text{mod}}}\) then maps each set of \(k_{\text{mod}}\) coded bits into a single complex symbol, producing the transmitted symbol vector \(\mathbf{x}\) of length
\[
n = \frac{N}{k_{\text{mod}}}.
\]
Hence, the overall rate of information transmission (in bits per complex transmission) is
\[
R = R_{\text{cod}}\, k_{\text{mod}}.
\]

\medskip

\cb{Such a two-step design does not necessarily provide the best performance, Instead, it is better to encode directly from information bits to complex valued codewords (combining binary coding and modulation). The goal of this paper is to propose a modular approach for realizing this coding using machine learning. Ideally, we want to increase the information rate $R$ at which we can achieve a desired FEP $\epsilon$ at a given SNR defined as
\[
SNR = \frac{P}{N_0}.
\]
Importantly, we would like our code to be testable, meaning that one can exhaustively test the code numerically with reasonable complexity. Note that the size of the codebook is $2^{nR}$ codewords, and exhaustive testing becomes quickly nonrealizable as $n$ or $R$ increases. 
Our proposed machine-learning based encoder/decoder takes this into account, as discussed next. }

%\textcolor{red}{Shannon's classical formula states that the capacity (in bits per transmission) is}
%\[
%C = \log_{2}(1 + \gamma).
%\]
%This capacity assumes infinite blocklength and a negligible error probability (\(\varepsilon \to 0\)), which is not representative of delay-constrained systems with short blocklengths.

% \subsection*{Neural Network-Based Joint Design}

% To improve performance at shorter blocklengths, we propose a joint neural-network-based design of the entire transmit, receive chain (encoder, modulator, demodulator, and decoder). \textcolor{red}{(define block and size)In particular, we implement multi-level coding and modulation for each 16-bit sized block \(m_{i}, i=1,\dots,N\) bits , where \(K = 16 \times N\).The complex block length is set to $n=64$ bits, so the rate is
% \[
% R = \frac{ 16 \times N}{n}.
% \]}
% Our approach jointly learns both the mapping of bits to symbols and the inverse mapping of symbols back to bits, constrained by an average power requirement. By training these neural networks in an end-to-end fashion, we can potentially approach the capacity region even at small \(n\), maintaining low latency while keeping the frame error probability \(\varepsilon\) below the target threshold. This stands in contrast to the conventional separated design of channel coding and modulation, offering a more holistic optimization for short-packet or delay-sensitive applications.

\section{Multi-Level Autoencoder (MLAE)}

In this section, we present our proposed solution, a multi-level autoencoder (MLAE) framework that integrates multi-level coding, deep neural network design, and an end-to-end training methodology.  The model is discussed in details in the following three subsections, highlighting: (1) how multi-level coding is realized by multiple encoder/decoder pairs, (2) the underlying neural network architecture of these pairs, and (3) the overall methodology, including training procedures and custom loss functions.

\subsection{Multi-Level Coding AE Design}
\begin{figure}
    \centering
    %\includesvg[width=0.5\textwidth, inkscapelatex=true]{deep_communication_final.svg}
    \includegraphics[width=0.5\textwidth]{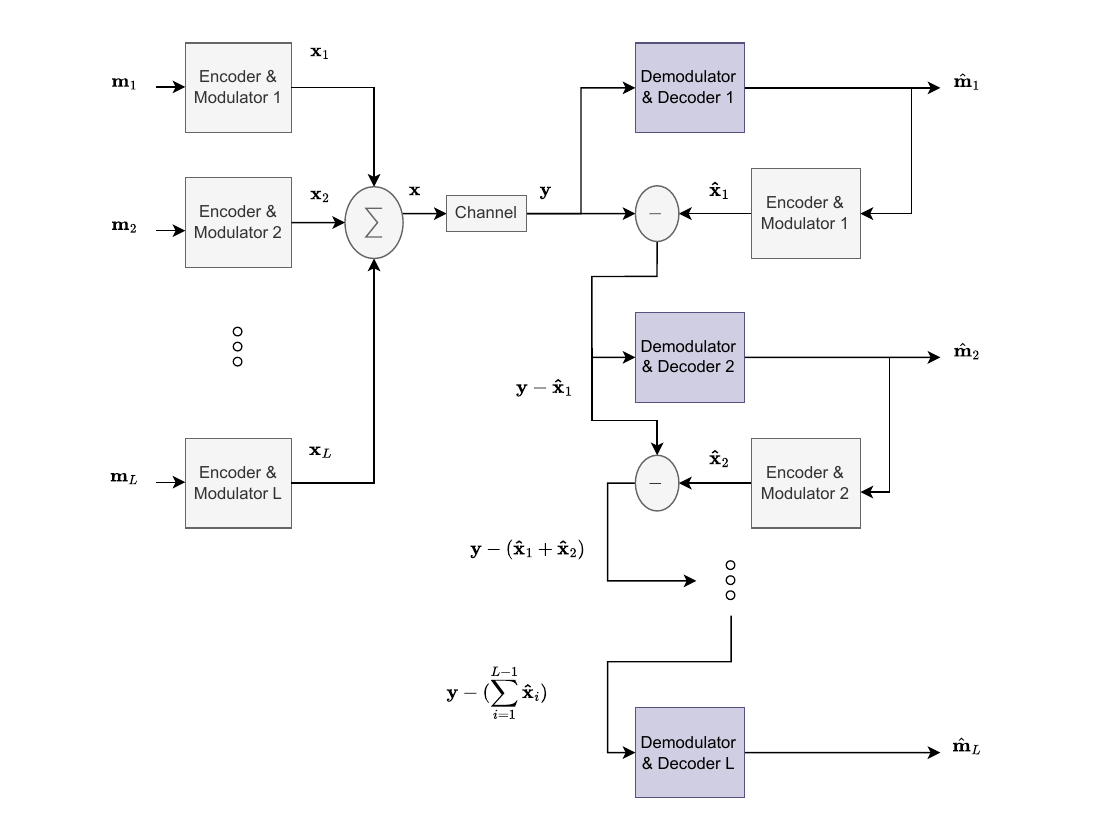}
   \caption{
    Detailed illustration of the proposed MLEA architecture. The input bit sequence $m_i,  i=1,\dots,L$ is first transformed by the encoder and modulator into \cb{a sub-codeword of length $n$. The output of all the $L$ encoders and modulators is added together then transmitted through a communication channel. Then, the MLAE framework adopts a successive decoding strategy, where each decoder begins with the current residual signal calculated by subtracting the contributions of already-decoded blocks from the received signal, and retrieves its assigned bits, re-encodes them, and subtracts this re-encoded estimate from the residual}.}
    \label{fig_mlae}
\end{figure}

The MLAE framework partitions the end-to-end coding chain into \cb{\(L\) distinct \emph{levels}, as illustrated in Figure~\ref{fig_mlae}. The input message \(m\) is divided into \(L\) sub-messages, \(m_{i}\) (\(i = 1, \dots, L\)), each consisting of $B$ bits. Thus, the total number of input bits is \(K = B \times L\). The block length is fixed at \(n\) complex symbols, resulting in a coding rate of \(R = \frac{B \times L}{n}\)}. Each level is tasked with independently encoding and modulating its input bits into a \cb{complex-valued vector, and the overall architecture combines all such vectors into a transmit signal. Upon transmitting this over a noisy channel, the decoder uses the multi-level structure to decode the bits for each level from the received signal.} The MLAE jointly learns both the mapping of bits to symbols and the inverse mapping of symbols back to bits, as detailed below.

\begin{itemize}
    \item \textbf{Multiple Input Levels:} 
    The model consist of $L$ levels. Each level receives a equal portion of the input\cb{ bits. Thus, the total number of input bits is $K = B \times L$.}

    \item \textbf{Encoders:} 
    Each level has a dedicated encoder network that outputs a 2D output of shape \(\texttt{(blocklength, 2)}\), where the second dimension represents the real and imaginary parts. These encoded signals are all summed together, which yields the transmit signal.

    \item \textbf{Decoders:}
    The \cb{summed transmit signals} passes through an additive white Gaussian noise (AWGN) channel. Each decoder level then attempts to reconstruct its corresponding bits. Instead of decoding all levels simultaneously from the \cb{received signal}, our MLAE framework decodes \cb{levels successively following a multi-stage decoding procedure \cite{intro_18}.} Concretely, each decoder begins with the current ``residual'' signal (the remaining portion of the received signal after subtracting out the already-decoded blocks). It recovers its assigned bits, re-encodes them, and subtracts this estimate from the residual. This subtraction step effectively removes that level’s contribution to the received signal, leaving a refined residual for the next decoder in the chain. The process repeats level by level, iteratively ``peeling off'' layers of interference and reducing the decoding burden for subsequent blocks. \cb{The advantage of this architecture is that it can be tested extensively, since each level can be tested separately while being able to provide performance guarantees for the whole architecture as we shall discuss later.} 

    %\item \textbf{Power Constraints:} 
    % To satisfy an average power constraint, the code applies normalizations (e.g., \(\texttt{normalize\_mod\_symbols\_tf}\)) to keep the transmit symbols near the desired power budget. Additionally, we define custom penalties for exceeding power limits.

\end{itemize}

Together, these elements enable the MLAE to map multiple input blocks onto a shared channel resource and iteratively decode each level’s data.

\subsection{Neural Network Design}

The MLAE architecture in our work combines multiple convolutional autoencoders into a multi-level system, where each level comprises a dedicated encoder--decoder pair. Each encoder employs several stacked one-dimensional (1D) convolutional layers with ELU activation functions and batch normalization to process the incoming bit-stream. After progressively refining the features in these convolutional layers, the output is reshaped and passed through a final 1D convolution to produce symbols in two real dimensions (in-phase and quadrature), whose size matches the \cb{desired blocklength. During training, symbols are normalized to maintain power constraints, ensuring that the average transmitted energy remains within a desired limit.} A typical configuration for one encoder is summarized in Table~\ref{tab:enc_arch}.

To model the noisy communication medium, we insert an additive white Gaussian noise (AWGN) layer between the encoder \cb{and decoder. On the receiving side,} each decoder mirrors the encoder by reversing these transformations. Several 1D convolutions with ReLU activations and batch normalization progressively map the complex symbols back to a probability distribution over bits, culminating in a sigmoid output layer to produce estimates of the transmitted binary sequence. Table~\ref{tab:dec_arch} illustrates the layer structure for a single decoder level. By stacking multiple such encoder--decoder pairs, we create a multi-level coding chain that superimposes and then iteratively separates distinct bit-stream \cb{blocks as shown in Fig. \ref{fig_mlae}.} 

\begin{table}[h]
    \centering
    \caption{\cb{Example encoder architecture for one level. Here the number of input bits to each level is $16$, and the codelength is $n=64$.}}
    \label{tab:enc_arch}
    \begin{tabular}{l|l|l|l|l}
    \hline
    \textbf{Layer} & \textbf{Filters} & \textbf{Kernel} & \textbf{Activation} & \textbf{Output Shape} \\
    \hline
    Input & - & - & - & $(16,1)$ \\
    1D Convolution & 200 & 2 & ReLU & $(16, 200)$ \\
    Batch Normalization & - & - & - & $(16, 200)$ \\
    1D Convolution & 200 & 2 & ReLU & $(16, 200)$ \\
    Batch Normalization & - & - & - & $(16, 200)$ \\
    1D Convolution & $n/4$ & 1 & ReLU & $(16,n/4)$ \\
    Reshape & - & - & - & $(n, 4)$ \\
    1D Convolution & 150 & 2 & ReLU & $(n, 150)$ \\
    Batch Normalization & - & - & - & $(n, 150)$ \\
    1D Convolution & 2 & 1 & - & $(n, 2)$ \\
    \hline
    \end{tabular}
\end{table}

\begin{table}[h]
    \centering
    \caption{\cb{Example decoder architecture for one level. Here, the number of information bits per level is $16$, and the codelength is $n=64$.}}
    \label{tab:dec_arch}
    \begin{tabular}{l|l|l|l|l}
    \hline
    \textbf{Layer} & \textbf{Filters} & \textbf{Kernel} & \textbf{Activation} & \textbf{Output Shape} \\
    \hline
    1D Convolution & 150 & 2 & ReLU & $(n, 150)$ \\
    Batch Normalization & - & - & - & $(n, 150)$ \\
    1D Convolution & 150 & 1 & - & $(n, 150)$ \\
    Batch Normalization & - & - & - & $(n, 150)$ \\
    1D Convolution & 200 & 2 & ReLU & $(n, 200)$ \\
    Batch Normalization & - & - & - & $(n, 200)$ \\
    1D Convolution & 200 & 2 & ReLU & $(n, 200)$ \\
    Batch Normalization & - & - & - & $(n, 200)$ \\
    1D Convolution & 1 & 4 & Sigmoid & $(16, 1)$ \\
    \hline
    \end{tabular}
\end{table}

\subsection{Training Methodology}

We trained the MLAE using a binary cross-entropy loss. Each decoder output corresponds to a reconstruction of its assigned bit-block, and a weighted sum of losses can be used to emphasize certain decoders during training. In early experiments, we explored a step-by-step procedure in which we trained one encoder--decoder pair at a time, freezing the other blocks until each pair achieved a reasonable accuracy. This stagewise approach helped each level to learn basic encoding and decoding features in isolation. However, to fully leverage the end-to-end nature of our design, we performed subsequent \emph{joint training} with all encoders and decoders unfrozen, which yielded higher overall performance and improved robustness against inter-level interference.

\cb{The training data consisted of all possible binary combinations for the target bit level, ensuring coverage of the full input space. for example, with $16$ bits as an input to a specific level, each codeword was repeated $3$ times with different noise realizations.} We used Adam optimizer \cite{adam} with a carefully tuned learning-rate schedule and employed standard strategies such as early stopping. In evaluation, we measured bit error rate (BER) over all possible binary combinations in the codebook for $2^{10}$ different noise realization per \cb{codeword, in other word, each possible binary combination was repeated $2^{10}$ times in the testing set to ensure exhaustive testing.} This insures the system’s reliability.

All blocks are trained jointly using a weighted cross-entropy loss, with a global power constraint applied to the transmitted symbols. In this approach the model learns optimal power allocation per level as part of the training.

% We explore two training methods:
% \begin{enumerate}
%     \item \textbf{Method 1:} Each level is trained separately by freezing the weights of all other levels and assigning a power constraint to each level based on expected \cb{noise power (interference from other layers).} This method proves difficult, particularly when handling a large number of blocks, and is highly sensitive to power constraints.
    
%     \item \textbf{Method 2:} All blocks are trained jointly using a weighted cross-entropy loss, with a global power constraint applied to the transmitted symbols. This approach is more stable since the model learns optimal power allocation as part of the training.
% \end{enumerate}

% \cb{Method 2 was used to train and evaluate the model performance.}
% 

% \begin{figure*}[!t]
%     \centering
%     \includesvg[width=1\textwidth, inkscapelatex=false]{MLAE_all.tex}
%     \caption{Comparison of BER versus rate (bits/transmission) for MLAE, TurboAE-MOD, and Polar codes under two different SNR conditions: (a) \( \text{SNR} = 0 \, \text{dB} \) and (b) \( \text{SNR} = 2.5 \, \text{dB} \).}
%     \label{fig_main_res}
% \end{figure*}

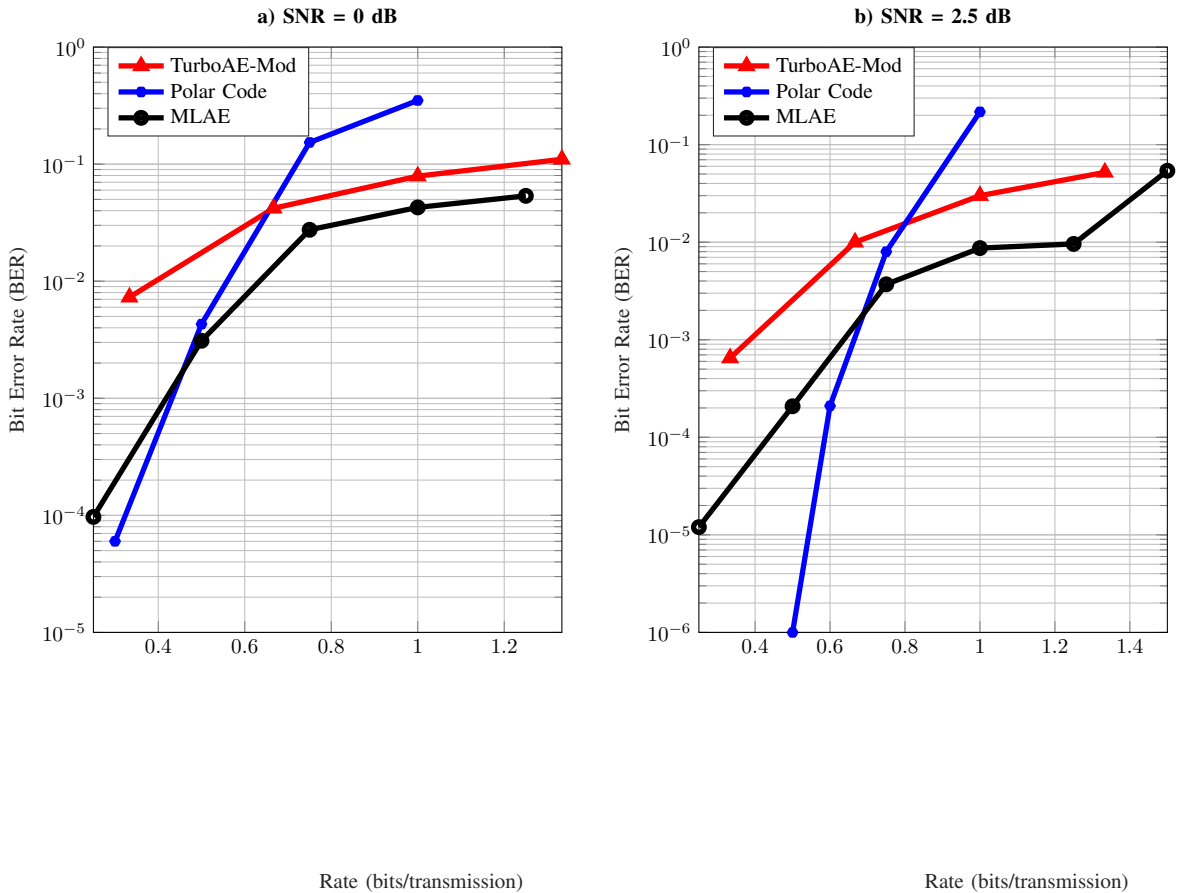
\begin{figure*}[t]
    \begin{center}
        \tikzset{every picture/.style={scale=1}, every node/.style={scale=0.8}}
\begin{tikzpicture}

\begin{axis}[%
width=2.452in,
height=3.066in,
at={(0.158in,0.481in)},
scale only axis,
xmin=0.25,
xmax=1.33333333,
xminorticks=true,
xlabel style={at={(axis description cs:0.7,-0.4)}, font=\color{white!15!black}},
xlabel={Rate (bits/transmission)},
ymode=log,
ymin=1e-05,
ymax=1,
yminorticks=true,
ylabel style={font=\color{white!15!black}},
ylabel={Bit Error Rate (BER)},
axis background/.style={fill=white},
title style={font=\bfseries},
title={a) SNR = 0 dB},
xmajorgrids,
xminorgrids,
ymajorgrids,
yminorgrids,
legend style={at={(0.03,1)}, anchor=north west, legend cell align=left, align=left, draw=white!15!black}
]
\addplot [color=red, line width=2.0pt, mark=triangle, mark options={solid, red}]
  table[row sep=crcr]{%
0.3333	0.0073\\
0.666666	0.042\\
1	0.079\\
1.33333333	0.11\\
};
\addlegendentry{TurboAE-Mod}

\addplot [color=blue, line width=2.0pt, mark=asterisk, mark options={solid, blue}]
  table[row sep=crcr]{%
0.25	0\\
0.3	6e-05\\
0.5	0.0043\\
0.75	0.1532\\
1	0.35\\
};
\addlegendentry{Polar Code}

\addplot [color=black, line width=2.0pt, mark=o, mark options={solid, black}]
  table[row sep=crcr]{%
0.25	9.7e-05\\
0.5	0.0031\\
0.75	0.0275\\
1	0.0427\\
1.25  0.0536 \\
};
\addlegendentry{MLAE}

\end{axis}

\begin{axis}[%
width=2.452in,
height=3.066in,
at={(3.327in,0.481in)},
scale only axis,
xmin=0.25,
xmax=1.5,
xminorticks=true,
xlabel style={at={(axis description cs:0.7,-0.4)},anchor=north,font=\color{white!15!black}},
xlabel={Rate (bits/transmission)},
ymode=log,
ymin=1e-06,
ymax=1,
yminorticks=true,
ylabel style={font=\color{white!15!black}},
ylabel={Bit Error Rate (BER)},
axis background/.style={fill=white},
title style={font=\bfseries},
title={b) SNR = 2.5 dB},
xmajorgrids,
xminorgrids,
ymajorgrids,
yminorgrids,
legend style={at={(0.03,1)}, anchor=north west, legend cell align=left, align=left, draw=white!15!black}
]
\addplot [color=red, line width=2.0pt, mark=triangle, mark options={solid, red}]
  table[row sep=crcr]{%
0.3333	0.00065\\
0.666666	0.01\\
1	0.03\\
1.33333333	0.052\\
};
\addlegendentry{TurboAE-Mod}

\addplot [color=blue, line width=2.0pt, mark=asterisk, mark options={solid, blue}]
  table[row sep=crcr]{%
0.25	0\\
0.5	1e-06\\
0.6	0.00021\\
0.75	0.008\\
1	0.2168\\
};
\addlegendentry{Polar Code}

\addplot [color=black, line width=2.0pt, mark=o, mark options={solid, black}]
  table[row sep=crcr]{%
0.25	1.2e-05\\
0.5	0.000208\\
0.75	0.0037\\
1	0.0087\\
1.25	0.0096\\
1.5     0.054 \\
};
\addlegendentry{MLAE}

\end{axis}

\begin{axis}[%
width=5.833in,
height=4.375in,
at={(0in,0in)},
scale only axis,
xmin=0,
xmax=1,
ymin=0,
ymax=1,
axis line style={draw=none},
ticks=none,
axis x line*=bottom,
axis y line*=left
]
\end{axis}
\end{tikzpicture}%
        \caption{Comparison of BER versus rate (bits/transmission) for MLAE, TurboAE-MOD, and Polar codes under two different SNR conditions: (a) \( \text{SNR} = 0 \, \text{dB} \) and (b) \( \text{SNR} = 2.5 \, \text{dB} \).}
        \label{fig_main_res}
    \end{center}
\end{figure*}

\section{Numerical Results}

In this section, we present the simulation results and compare the proposed approach with TurboAE-MOD~\cite{turboAE-mod} and Polar codes with BPSK modulation. For the Polar code implementation, we use polarization-adjusted convolutional (PAC) codes, as proposed by Arıkan, which introduce a one-to-one convolutional transform before the polar transform. List decoding is employed for the PAC codes, and they are optimized based on the technique described in \cite{polar_codes}.

\subsection{Simulation Setup}

We train and test our MLAE model under an additive white Gaussian noise (AWGN) channel at a blocklength of \cb{$n = 64$, where we chose the input of each level to be $16$ bits.} For training, we generate a dataset of $3 \times 2^{16}$ input sequences, covering all permutations of a 16-bit binary sequence and replicating each sequence three times to simulate different noise realizations\cb{. Although the number of repetitions (3) of each codeword is small,  we shall see later through exhaustive testing that the MLAE can learn to encode/decode reliably nonetheless.}  For testing, we employ a large set of $2^{10} \times 2^{16}$ sequences per level, encompassing all possible codewords in the \cb{codebook for a level} transmitted over $2^{10}$ different noise realizations. \cb{This exhaustive testing provides confidence in the evaluated performance of our design. Note that comparatively conducting a similar test for the TurboAE-MOD would require testing $2^{64}$ possible sequences, which is prohibitive.}

\cb{Adam optimizer was used for training the model with an initial learning rate of 0.001, which is exponentially decayed. We use a batch size of 1024 and train for 100 epochs. The training SNR is fixed, and the model is evaluated at various SNR values by selectively removing one or more blocks to lower the coding rate without re-training the model, thereby reducing interference and enabling reliable transmission over noisier channels.}

\subsection{Simulation Results}

\cb{To compute the aggregate BER across multiple blocks, we calculate the BER for each level, then sum them and divide by $L$ to find the total BER.}

% \[
% \text{BER}_{total} = 1-\prod_{i=1}^{N}{(1-\text{BER}_i)}.
% \]

Table~\ref{tab:results_block} summarizes the BER for each individual level at $0$ dB SNR, as well as the combined \emph{Total} BER. 

\cb{Fig. \ref{fig_main_res} shows the BER vs transmission rate for MLAE, Polar code, and TurboAE-MOD on two different SNR values (0 dB and 2.5 dB).} Our MLAE model demonstrates strong performance across a range of rates and SNR settings, often matching or even exceeding TurboAE-MOD. While TurboAE-MOD still achieves lower BER in certain cases, MLAE is validated on a far more comprehensive test set covering all possible codewords in the codebook across $2^{10}$ noise realizations. This extensive evaluation \cb{provides high confidence in the simulation results.}
  
% Our proposed MLAE model outperforms Polar codes and achieves performance comparable to TurboAE-MOD. Notably, TurboAE-MOD shows discrepancies between its bit-error rate (BER) on the training set and on the testing set, whereas our model is validated on a far more comprehensive test set covering all possible codewords in the codebook across $2^{10}$ noise realizations. This extensive evaluation underscores the superior reliability of our approach.

%For the TurboAE-MOD, which maintains a constant coding rate of $R=1/3$, we fix the blocklength at 64 symbols by reducing the number of information bits accordingly. In contrast, MLAE encodes 16 information bits per block, resulting in a rate of $16/64$, and adjusts this rate by dynamically adding or removing blocks. This flexibility in coding rate, without the need to retrain the model, is another key strength of MLAE, allowing the system to adapt to various channel conditions efficiently.

%Moreover, we investigate multiple architectures, including feed-forward neural networks (FFNNs), recurrent neural networks (RNNs), transformers, and convolutional neural networks (CNNs), as well as several activation functions (ReLU, ELU, and GELU). Among these, CNNs deliver the best trade-off between complexity and performance, while ReLU offers superior results relative to ELU and GELU. These design choices highlight the robust and practical advantages of our model for modern communication systems.

\begin{table}[h]
    \centering
    \caption{MLAE ber results at $0$ \textit{dB} snr for different levels and coding rates. 
    The `Total' column shows the cumulative ber across all active levels.}
    \label{tab:results_block}
    \begin{tabular}{c|c|c|c|c|c}
    \hline
      & \multicolumn{5}{c}{\textbf{Bit Error Rate (BER)}} \\
    \cline{2-6}
    \textbf{Rate} 
      & \textbf{Level 1} 
      & \textbf{Level 2} 
      & \textbf{Level 3} 
      & \textbf{Level 4} 
      & \textbf{Total} \\
    \hline
    $0.25$ 
      & $9.7 \times 10^{-5}$ 
      & - 
      & - 
      & - 
      & $9.7 \times 10^{-5}$ \\
    \hline
    $0.5$ 
      & $0.0027$ 
      & $0.0035$ 
      & - 
      & - 
      & $0.0031$ \\
    \hline
    $0.75$ 
      & $0.008$ 
      & $0.074$ 
      & $0.0049$ 
      & - 
      & $0.0275$ \\
    \hline
    $1$ 
      & $0.01$ 
      & $0.11$ 
      & $0.022$ 
      & $0.02$ 
      & $0.0427$ \\
    \hline
    \end{tabular}
\end{table}

\section{CONCLUSION}
In this paper, we introduced a multi-level convolutional autoencoder (MLAE) design for coding and modulation over Gaussian channels. By splitting the message into $B$-bit blocks and assigning a dedicated encoder–decoder pair to each level, the proposed system can systematically cover the entire \cb{codebook during training/testing, }offering a rigorous and practical benchmark that circumvents the sampling limitations of previous AI-based solutions. The reported results not only highlight the model’s ability to achieve comparable, or even superior performance to classical polar codes and Turbo-AE, but also underscore its adaptability; by simply discarding one or more blocks, the MLAE can tailor its coding rate to changing channel conditions without retraining. These attributes are especially relevant for next-generation wireless systems, where stringent latency, reliability, and scalability requirements coexist. Future research directions may include expanding MLAE to address more complex channel scenarios such as fading and interference. In addition, our experiments show that increasing the number of levels  $L$ to accommodate higher rates poses training challenges, primarily due to power allocation, indicating that further refinements in the training strategy are needed.

\bibliographystyle{IEEEtran}
\bibliography{references}

\end{document}